\documentclass[aps,prl,twocolumn]{revtex4}
\usepackage{graphicx}
\usepackage{amsmath}
\usepackage{amssymb}

\begin{document}

\title{Quasi-particle Lifetime in a Mixture of Bose and Fermi Superfluids }
\author{Wei Zheng and Hui Zhai}
\affiliation{Institute for Advanced Study, Tsinghua University, Beijing, 100084, China}
\date{\today }

\begin{abstract}
In this letter, to reveal the effect of quasi-particle interactions in a
Bose-Fermi superfluid mixture, we consider the lifetime of quasi-particle of
Bose superfluid due to its interaction with quasi-particles in Fermi
superfluid. We find that this damping rate, i.e. inverse of the lifetime,
has quite different threshold behavior at the BCS and the BEC side of the
Fermi superfluid. The damping rate is a constant nearby the threshold
momentum in the BCS side, while it increases rapidly in the BEC side. This
is because in the BCS side the decay processe is restricted by constant
density-of-state of fermion quasi-particle nearby Fermi surface, while such
a restriction does not exist in the BEC side where the damping process is
dominated by bosonic quasi-particles of Fermi superfluid. Our results are
related to collective mode experiment in recently realized Bose-Fermi
superfluid mixture.
\end{abstract}

\maketitle


Recently, for the first time, ENS group has realized a mixture of Bose and
Fermi superfluids \cite{ENS}. They prepare a mixture of bosonic ${}^7$Li
atoms and two spin components of fermionic ${}^6$Li atoms nearby an $s$-wave
Feshbach resonance between fermions. At low enough temperature, bosonic
atoms condense and become a Bose superfluid, while fermionic atoms form
pairs and become a Fermi superfluid. This experimental development generates
many interesting questions on interaction effects between these two types of
superfluid \cite{Stringari,ad}.

\begin{figure}[t]
\includegraphics[width=3.5 in]
{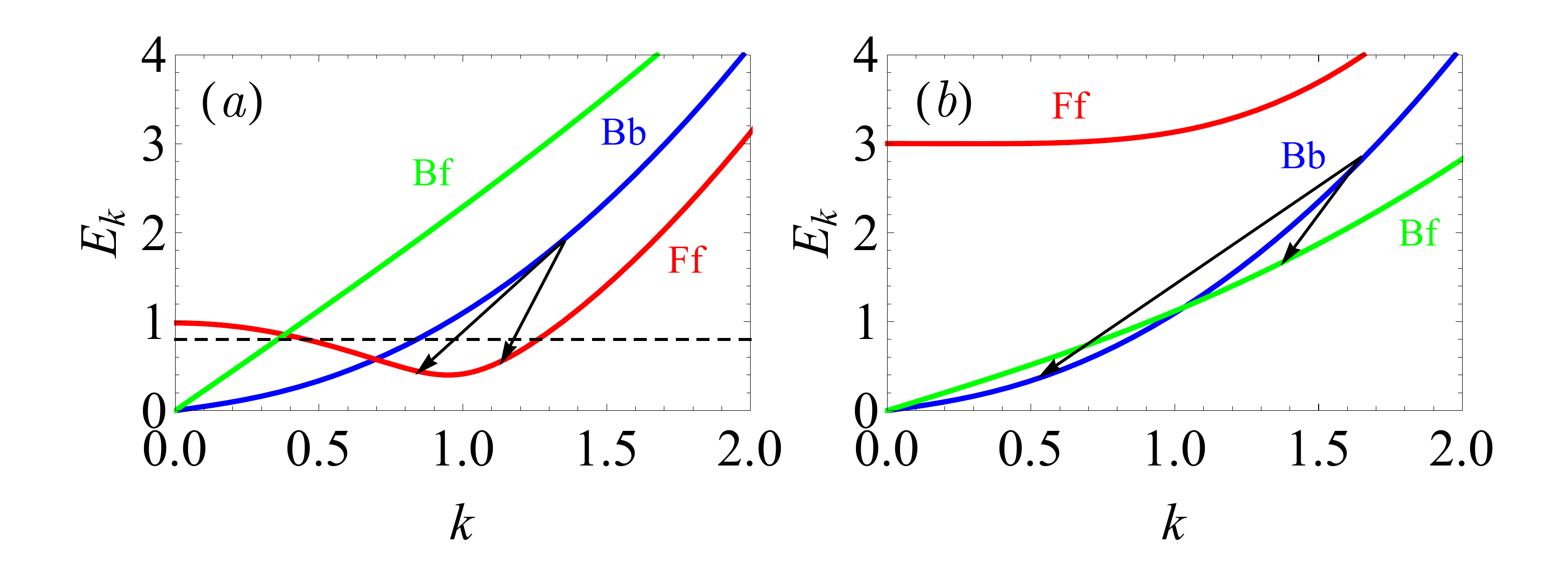}
\caption{Schematic of dispersions of bosonic mode of Bose superfluid ($B_{%
\text{b}}$), bosonic mode of Fermi superfluid ($B_{\text{f}}$) and fermionic
pair breaking mode of Fermi superfluid ($F_{\text{f}}$), respectively, at
the BCS side (a) and at the BEC side (b). In (a), dashed line represents
value of $2\Delta $. Arrows denote that $B_{\text{b}}$ mode decays into two $%
F_{\text{f}}$ modes. In (b), arrows denote that $B_{\text{b}}$ mode is
scattered by generating an additional $B_{\text{f}}$ mode. }
\label{BEC_BCS}
\end{figure}

Elementary excitations and their interactions play an important role in
quantum many-body system. Here we can compare the low-energy elementary
excitations of this superfluid mixture with other two widely studied
mixtures, i.e. mixture of a BEC with normal Fermi gas \cite{BF} and mixture
of two BECs \cite{BB}. A Bose-Fermi superfluid mixture exhibits two gapless
bosonic modes (denoted by $B_{\mathrm{b}}$ and $B_{\mathrm{f}}$ in Fig. \ref%
{BEC_BCS}), corresponding to Goldstone modes of Bose superfluidity ($B_{%
\mathrm{b}}$) and Fermi superfluidity ($B_{\mathrm{f}}$), respectively, and
a gapped fermionic excitation that describes the Cooper pair breaking
(denoted by $F_{\mathrm{f}}$ in Fig. \ref{BEC_BCS}). While in the mixture of
a BEC with normal Fermi gas, there exists only one bosonic Goldstone mode
and the fermionic excitation (particle or hole excitation) is always gapless
at the Fermi surface. Mixture of two BECs also exhibits two bosonic
Goldstone modes but there is no fermionic excitation in this system.

Moreover, in the cold atom system the Fermi superfluid can be continuously
tuned from the BCS regime to the BEC regime by utilizing the Feshbach
resonance. In the BCS limit, as schematically shown in Fig. \ref{BEC_BCS}%
(a), it is known that the $B_{\mathrm{f}}$ mode has a quite large velocity
proportional to $v_{\text{F}}/\sqrt{3}$ \cite{FGoldstone}, while the gap of
the $F_{\mathrm{f}}$ mode is exponentially small. As approaching the BEC
side, as shown in Fig. \ref{BEC_BCS}(b), the gap of the $F_{\mathrm{f}}$
mode becomes larger and larger, and on the other hand, the velocity of the $%
B_{\mathrm{f}}$ mode becomes smaller and smaller \cite{RMP}.

Therefore, the interplay between these three modes is quite unique in the
Bose-Fermi superfluid mixture, and it will lead to different behaviors in
the BCS and the BEC sides of Fermi superfluid. One manifestation of
interaction between elementary excitations is the lifetime of
quasi-particles. The most well-known effect is Landau-Beliaev damping in the
Bose superfluid \cite{Landau_Baliaev}. Interaction between bosonic mode
itself gives rise to a finite lifetime of the bosonic quasi-particle. The
damping rate, as the inverse of the lifetime $\gamma =1/\tau $, is
proportional to $k^{5}$ at zero-temperature and to $T^{4}$ at finite
temperature \cite{Pethick, Stringari_book}. This effect has been
experimentally studied in atomic BEC by measuring the damping rate of
collective modes \cite{exp} and theoretically works have also been carried
out in the content of cold atom systems \cite{Vincent, Stringari_2, Gora}.
Landau damping has also been studied for mixture of BEC with normal Fermi
gas \cite{dampmix} and dipolar BEC \cite{dampdipole}.

In this letter we present an alternative damping channel for bosonic
quasi-particle of Bose superfluid ($B_{\mathrm{b}}$ mode) due to its
interaction with quasi-particles in Fermi superfluid ($B_{\mathrm{f}}$ and $%
F_{\mathrm{f}}$ modes). We focus on the typical cold atom situation that
Fermi superfluid is in the strongly interacting regime while Bose superfluid
is in the weakly interacting regime. We show that this damping mechanism
will be activated only when momentum of $B_{\mathrm{b}}$ excitation exceeds
a critical value $k_{\text{c}}$. We investigate the threshold behavior of
damping rate $\gamma =\mathcal{C}(k-k_{\text{c}})^{\alpha }$, and the key
result is that we find different $\alpha $ for the BCS side and the BEC side
of Fermi superfluid.

\textit{Model.} We consider a mixture of bosons and spin-$1/2$ fermions,
whose Hamiltonian is given by
\begin{align}
& \hat{H}_{\text{f}}=\int d^{3}\mathbf{r}\left\{ \hat{c}_{\sigma }^{\dag }(%
\mathbf{r})H_{0,\text{f}}\hat{c}_{\sigma }(\mathbf{r})-g_{\text{f}}\hat{c}%
_{\uparrow }^{\dag }(\mathbf{r})\hat{c}_{\downarrow }^{\dag }(\mathbf{r})%
\hat{c}_{\downarrow }(\mathbf{r})\hat{c}_{\uparrow }(\mathbf{r})\right\}
\notag \\
& \hat{H}_{\text{b}}=\int d^{3}\mathbf{r}\left\{ \hat{b}^{\dag }(\mathbf{r}%
)H_{0,\text{b}}\hat{b}(\mathbf{r})+\frac{g_{\text{b}}}{2}\hat{b}^{\dag }(%
\mathbf{r})\hat{b}^{\dag }(\mathbf{r})\hat{b}(\mathbf{r})\hat{b}(\mathbf{r}%
)\right\}  \notag \\
& \hat{H}_{\text{bf}}=g_{\text{bf}}\int d^{3}\mathbf{r}\hat{b}^{\dag }(%
\mathbf{r})\hat{b}(\mathbf{r})\hat{c}_{\sigma }^{\dag }(\mathbf{r})\hat{c}%
_{\sigma }(\mathbf{r})
\end{align}%
where $H_{0,i}=-\hbar ^{2}\nabla ^{2}/(2m_{i})-\mu _{i}$ and $i=\text{b},%
\text{f}$ denotes bosons or fermions. Since interaction between fermions is
nearby a Feshbach resonance, we shall relate $g_{\text{f}}$ to scattering
length $a_{\text{f}}$ as $1/g_{\text{f}}=m/(4\pi \hbar ^{2}a_{\text{f}%
})+\sum_{\mathbf{k}}1/\left[ 2\epsilon _{\text{f}}(\mathbf{k})\right] $ with
$\epsilon _{\text{f}}(\mathbf{k})=\hbar ^{2}\mathbf{k}^{2}/(2m_{\text{f}})$.
The ground state of $\hat{H}_{\text{f}}$ is a superfluid of fermion pairs.
Applying the BCS-BEC crossover mean-field theory to $\hat{H}_{\text{f}}$ one
can obtain a gapped fermion $F_{\text{f}}$ mode with excitation energy $%
\mathcal{E}_{F_{\text{f}}}=\sqrt{[\epsilon _{\text{f}}(\mathbf{k})-\mu _{%
\text{f}}]^{2}+\Delta ^{2}}$. As $-1/(k_{\text{F}}a_{\text{f}})$ decreases
from the BCS side to the BEC side, $\mu _{\text{f}}$ decreases and $\Delta $
increases \cite{RMP}. $\hat{H}_{\text{f}}$ also has a bosonic $B_{\text{f}}$
mode that describes center-of-mass motion of Cooper pairs, which has a
phonon-like dispersion $\mathcal{E}_{B_{\text{f}}}=\hbar c_{\text{f}}k$, and
$c_{\text{f}}$ evolves smoothly from $v_{\text{F}}/\sqrt{3}$ to $\sqrt{\pi
\hbar ^{2}a_{\text{m}}n_{\text{m}}/m_{\text{f}}^{2}}$ \cite{RMP,Thomas},
where $a_{\text{m}}=0.6a_{\text{f}}$ is the scattering length between
fermion pairs \cite{Petrov} and $n_{\text{m}}$ is molecule density. For
equal population case $n_{\text{m}}=n_{\uparrow }=n_{\downarrow }=n_{\text{f}%
}$.

When magnetic field locates nearby a Feshbach resonance between fermions,
generically $g_{\text{b}}$ and $g_{\text{bf}}$ terms are in the weakly
interacting regime and can be treated by Bogoliubov approximation. In the
leading order of $n_{\text{b}}$ ($n_{\text{b}}=N_{\text{b}}/V$, $N_{\text{b}%
} $ is condensate bosonic atoms), we replace two of $\hat{b}^{\dag }$ or $%
\hat{b}$ operator with $\sqrt{N_{\text{b}}}$ in the interaction part. From $%
\hat{H}_{\text{b}}$ we obtain a Bogoliubov spectrum for Bose superfluid $%
\mathcal{E}_{B_{\text{b}}}=\sqrt{\epsilon _{\text{b}}(\mathbf{k})[\epsilon _{%
\text{b}}(\mathbf{k})+2g_{\text{b}}n_{\text{b}}]}$, where $\epsilon _{\text{b%
}}(\mathbf{k})=\hbar ^{2}\mathbf{k}^{2}/(2m_{\text{b}})$ and $g_{\text{b}%
}=4\pi \hbar ^{2}a_{\text{b}}/m_{\text{b}}$. When $k\ll 1/\xi =\sqrt{8\pi a_{%
\text{b}}n_{\text{b}}}$, the excitation is in the phonon regime with a
linear dispersion $\hbar c_{\text{b}}k$ and $c_{\text{b}}=\sqrt{4\pi \hbar
^{2}a_{\text{b}}n_{\text{b}}/m_{\text{b}}^{2}}$. When $k\gg 1/\xi =\sqrt{%
8\pi a_{\text{b}}n_{\text{b}}}$, the excitation is in the free-particle
regime with a quadratic dispersion $\epsilon _{\text{b}}(\mathbf{k})+g_{%
\text{b}}n_{\text{b}}$. Also in the leading order, $\hat{H}_{\text{bf}}$
becomes $g_{\text{bf}}n_{\text{b}}\int d^{3}\mathbf{r}c_{\sigma }^{\dag }(%
\mathbf{r})c_{\sigma }(\mathbf{r})$, which simply provides a constant shift
of chemical potential and will not affect spectrum and wave function of
quasi-particles. In the sub-leading order of $n_{\text{b}}$, only one $\hat{b%
}^{\dag }$ or $\hat{b}$ operator is replaced by $\sqrt{N_{\text{b}}}$, and
it describes interaction between quasi-particles. In this order, $\hat{H}_{%
\text{b}}$ leads to Landau-Beliaev damping discussed before \cite%
{Pethick,Stringari_book}. As we will show later, $\hat{H}_{\text{bf}}$ gives
rise to interaction between quasi-particles of Bose superfluid and those of
Fermi superfluid.

\begin{figure}[t]
\includegraphics[width=3.4 in]
{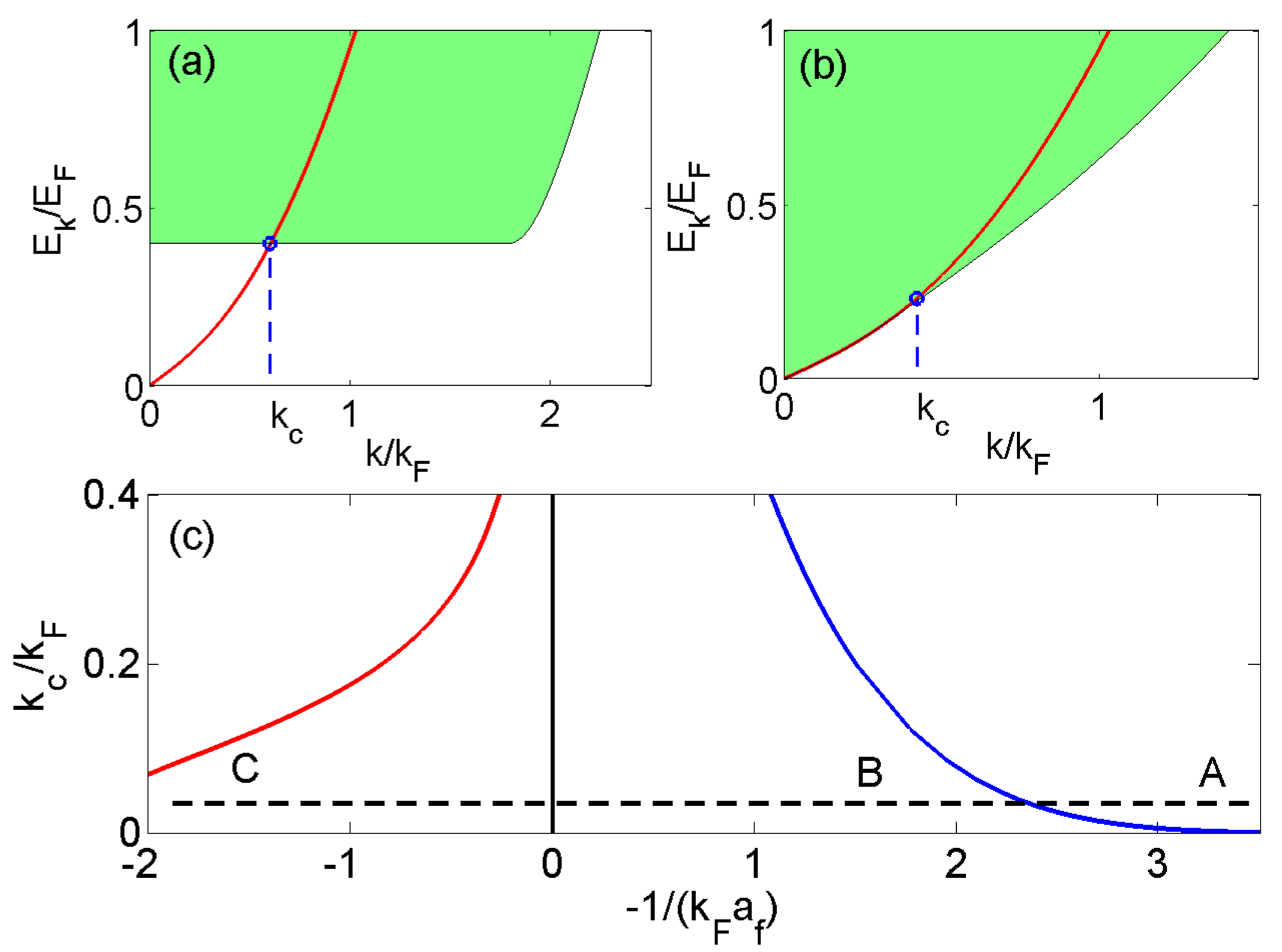}
\caption{ (a-b): Shaded area is a schematic of two-particle continuum for
two different damping channels, $F_{\text{f}}+F_{\text{f}}$ for (a) and $B_{%
\text{f}}+B_{\text{b}}$ for (b), corresponding to processes illustrated in
Fig. \protect\ref{BEC_BCS} (a) and (b), respectively. The red solid line is
dispersion of bosonic quasi-particle $B_{\text{b}}$ of Bose superfluid. $k_{%
\text{c}}$ marks the threshold momentum. (c) $k_{\text{c}}/k_{\text{F}}$ as
a function of $-1/(k_{\text{F}}a_{\text{f}})$. A, B and C mark three typical
regimes discussed in text. Below (above) the dashed line $k_{\text{c}}$ is
in the phonon (free-particle) regime of Bogoliubov dispersion for $B_{\text{b%
}}$ mode. For A and B, $k_{\text{c}}$ is given by (a), and $k_{\text{c}}$ is
in phonon regime for (A) and is in free-particle regime for (B). For (C), $%
k_{\text{c}}$ is given by (b) and is in free-particle regime. }
\label{Threshold}
\end{figure}

\textit{Damping Threshold.} There are two different decay channels for
bosonic quasi-particle $B_{\text{b}}$ of Bose superfluid. The first is decay
into two fermionic quasi-particles $F_{\text{f}}$ of Fermi superfluid, i.e. $%
B_{\text{b}}(\mathbf{k})\rightarrow F_{\text{f}}(\mathbf{k-q})+F_{\text{f}}(%
\mathbf{q})$, as shown in Fig. \ref{BEC_BCS}(a). In this case, the
energy-momentum conservation requires $\mathcal{E}_{B_{\text{b}}}(\mathbf{k}%
)=\mathcal{E}_{F_{\text{f}}}(\mathbf{k-q})+\mathcal{E}_{F_{\text{f}}}(%
\mathbf{q})$. Since $\mathcal{E}_{F_{\text{f}}}$ is gapped and the minimum
of $\mathcal{E}_{F_{\text{f}}}(\mathbf{k})$ is $\Delta $ occurring at $k_{0}$
with $k_{0}=\sqrt{2m_{\text{f}}\mu _{\text{f}}/\hbar ^{2}}$ for $\mu _{\text{%
f}}>0$ and $k_{0}=0$ for $\mu _{\text{f}}<0$, a typical two-particle
continuum for two $F_{\text{f}}$ modes in the BCS side is shown in Fig. \ref%
{Threshold}(a), which has a minimum of $2\Delta $ for $k<2k_{0}$. For this
channel, $k_{\text{c}}$ is determined by $\mathcal{E}_{B_{\text{b}}}$
meeting this two-particle threshold. In the BCS side of resonance, $k_{\text{%
c}}$ can be determined by equation $\mathcal{E}_{B_{\text{b}}}(k_{\text{c}%
})=2\Delta $ as long as the solution of $k_{\text{c}}$ is smaller than $%
2k_{0}$. Therefore, as $-1/(k_{\text{F}}a_{\text{f}})$ decreases from the
BCS side to unitary regime, $k_{\text{c}}$ increases as shown in Fig. \ref%
{Threshold}(c). Moreover, when $\Delta \ll \hbar ^{2}/(m_{\text{b}}\xi ^{2})$%
, $k_{\text{c}}$ is in the phonon regime of $B_{\text{b}}$ mode, while on
contrary, when $\Delta \gg \hbar ^{2}/(m_{\text{b}}\xi ^{2})$, $k_{\text{c}}$
is in the free-particle regime of $B_{\text{b}}$ mode.

The second channel is decay into two bosonic quasi-particles $B_{\text{f}}$,
i.e. $B_{\text{b}}(\mathbf{k})\rightarrow B_{\text{f}}(\mathbf{k-q})+B_{%
\text{f}}(\mathbf{q})$, or one $B_{\text{f}}$ and one $B_{\text{b}}$, i.e. $%
B_{\text{b}}(\mathbf{k})\rightarrow B_{\text{b}}(\mathbf{k-q})+B_{\text{f}}(%
\mathbf{q})$, as shown in Fig. \ref{BEC_BCS}(b). Since in the strongly
interacting regime of Fermi superfluid, $c_{\text{f}}$ is usually much
larger than $c_{\text{b}}$ because $a_{\text{m}}=0.6a_{\text{f}}\gg a_{\text{%
b}}$, it is easy to show that the two-particle threshold of $B_{\text{b}}+B_{%
\text{f}}$ is always lower than that of $B_{\text{f}}+B_{\text{f}}$. It is
also straightforward to show that $\mathcal{E}_{B_{\text{b}}}(\mathbf{k})$
coincides with two-particle threshold of $B_{\text{b}}+B_{\text{f}}$ up to $%
k_{\text{c}}$, as shown in Fig. \ref{Threshold}(b). That means for $k<k_{%
\text{c}}$, only the process with $\mathbf{q}=0$ can happen which in fact
does not lead to decay of quasi-particle. Thus, damping will be activated
only when $\mathcal{E}_{B_{\text{b}}}(\mathbf{k})$ is above two-particle
threshold when $k>k_{\text{c}}$, and $k_{\text{c}}$ is determined by $%
\partial \mathcal{E}_{B_{\text{b}}}(k)/\partial \left( \hbar k\right)
|_{k=k_{\text{c}}}=c_{\text{f}}$. Also due to $c_{\text{f}}\gg c_{\text{b}}$%
, $k_{\text{c}}$ is always located in the free-particle regime of $B_{\text{b%
}}$ mode.

Hence our following discussion can be divided into three representative
cases, as shown in Fig. \ref{Threshold}(c): Case A and B are both at the BCS
side of Fermi superfluid, where damping is determined by the first process.
For Case A, $\Delta \ll \hbar ^{2}/(m_{\text{b}}\xi ^{2})$ and therefore $k_{%
\text{c}}$ is in the phonon regime of $B_{\text{b}}$ mode. For Case B, $%
\Delta \gg \hbar ^{2}/(m_{\text{b}}\xi ^{2})$, and thus $k_{\text{c}}$ is in
the free-particle regime of $B_{\text{b}}$ mode. Case C is at the BEC side
of Fermi superfluid, where damping is determined by the second process, and $%
k_{\text{c}}$ is in the free-particle regime of $B_{\text{b}}$ mode.

\textit{Case A.} In this regime we start with BCS mean-field Hamiltonian for
$\hat{H}_{\text{f}}$ and Bogoliubov Hamiltonian for $\hat{H}_{\text{b}}$
given by
\begin{align}
& \hat{H}_{\text{f}}=\sum_{\mathbf{k}}\mathcal{E}_{F_{f}}(\mathbf{k})(\hat{%
\beta}_{\mathbf{k}}^{\dag }\hat{\beta}_{\mathbf{k}}+\hat{\gamma}_{\mathbf{k}%
}^{\dag }\hat{\gamma}_{\mathbf{k}}), \\
& \hat{H}_{\text{b}}=\sum_{\mathbf{k}}\mathcal{E}_{B_{b}}(\mathbf{k})\hat{%
\alpha}_{\mathbf{k}}^{\dag }\hat{\alpha}_{\mathbf{k}}
\end{align}%
where quasi-particle $\hat{\alpha}_{\mathbf{k}}$, $\hat{\beta}_{\mathbf{k}}$
and $\hat{\gamma}_{\mathbf{k}}$ are related to $\hat{b}_{\mathbf{k}}$ and $%
c_{\mathbf{k}\sigma }$ via $\hat{b}_{\mathbf{k}}=u_{\mathbf{k}}^{\text{b}}%
\hat{\alpha}_{\mathbf{k}}-v_{\mathbf{k}}^{\text{b}}\hat{\alpha}_{\mathbf{-k}%
}^{\dag }$, $\hat{c}_{\mathbf{k}\uparrow }=u_{\mathbf{k}}^{\text{f}}\hat{%
\beta}_{\mathbf{k}}+v_{\mathbf{k}}^{\text{f}}\hat{\gamma}_{\mathbf{-k}%
}^{\dag }$ and $\hat{c}_{\mathbf{k}\downarrow }=u_{\mathbf{k}}^{\text{f}}%
\hat{\gamma}_{\mathbf{k}}-v_{\mathbf{k}}^{\text{f}}\hat{\beta}_{\mathbf{-k}%
}^{\dag }$. Here $u_{\mathbf{k}}^{\text{b}}(v_{\mathbf{k}}^{\text{b}})=\sqrt{%
\frac{1}{2}\left( \frac{\epsilon _{\text{b}}(\mathbf{k})+g_{\text{b}}n_{%
\text{b}}}{\mathcal{E}_{B_{\text{b}}}(\mathbf{k})}\pm 1\right) }$ and $u_{%
\mathbf{k}}^{\text{f}}(v_{\mathbf{k}}^{\text{f}})=\sqrt{\frac{1}{2}\left(
1\pm \frac{\epsilon _{\text{f}}(\mathbf{k})-\mu _{\text{f}}}{\mathcal{E}_{F_{%
\text{f}}}(\mathbf{k})}\right) }$.

Now we discuss $\hat{H}_{\text{bf}}$ in the order of $\sqrt{n_{\text{b}}}$
by replacing one of $\hat{b}$ or $\hat{b}^{\dag }$ operator as $\sqrt{N_{%
\text{b}}}$, which leads to
\begin{equation}
\hat{H}_{\text{bf}}=g_{\text{bf}}\sqrt{\frac{n_{\text{b}}}{V}}\sum\limits_{%
\mathbf{k}\mathbf{q}}(\hat{c}_{\mathbf{k+q},\sigma }^{\dag }\hat{c}_{\mathbf{%
q},\sigma }\hat{b}_{\mathbf{k}}+\text{h.c.}).
\end{equation}%
We can further rewrite $\hat{H}_{\text{bf}}$ in term of quasi-particle
operators $\hat{\alpha}$, $\hat{\beta}$ and $\hat{\gamma}$. Here we focus on
zero-temperature damping rate (or lifetime) of bosonic $\alpha $ mode, thus,
only one term retains as \cite{supple}
\begin{align}
& g_{\text{bf}}\sqrt{\frac{n_{\text{b}}}{V}}\sum\limits_{\mathbf{k}\mathbf{q}%
}\mathcal{M}_{\mathbf{k}\mathbf{q}}\hat{\beta}_{\mathbf{k-q}}^{\dag }\hat{%
\gamma}_{\mathbf{q}}^{\dag }\hat{\alpha}_{\mathbf{k}}, \\
& \mathcal{M}_{\mathbf{k}\mathbf{q}}=(u_{\mathbf{k}}^{\text{b}}-v_{\mathbf{k}%
}^{\text{b}})(u_{\mathbf{k-q}}^{\text{f}}v_{\mathbf{q}}^{\text{f}}+v_{%
\mathbf{k-q}}^{\text{f}}u_{\mathbf{q}}^{\text{f}})
\end{align}%
This term describes the process that one $B_{\text{b}}$ mode decays into two
$F_{\text{f}}$ modes, as schematically drawn in Fig. \ref{BEC_BCS}(a). With
Fermi-Golden rule, the damping rate is given by
\begin{equation}
\gamma (\mathbf{k})=\frac{2\pi }{\hbar }\frac{n_{\text{b}}g_{\text{bf}}^{2}}{%
V}\sum_{\mathbf{q}}|\mathcal{M}_{\mathbf{k}\mathbf{q}}|^{2}\delta \left[
\mathcal{E}_{F_{\text{f}}}(\mathbf{k-q})+\mathcal{E}_{F_{\text{f}}}(\mathbf{q%
})-\mathcal{E}_{B_{\text{b}}}(\mathbf{k})\right] .  \label{rate_BCS}
\end{equation}

When $k_{\text{c}}$ is in the phonon regime, we can approximate $u_{\mathbf{k%
}}^{\text{b}}(v_{\mathbf{k}}^{\text{b}})=\sqrt{\frac{g_{\text{b}}n_{\text{b}}%
}{2\hbar c_{\text{b}}k}}\pm \frac{1}{2}\sqrt{\frac{\hbar c_{\text{b}}k}{2g_{%
\text{b}}n_{\text{b}}}}$. And since the decay products of $F_{\text{f}}$
mode locate nearby its minimum of dispersion $\mathcal{E}_{F_{\text{f}}}(%
\mathbf{k})$ at $k_{0}$, to the leading order we can approximate $u_{\mathbf{%
k}}^{\text{f}}(v_{\mathbf{k}}^{\text{f}})=1/\sqrt{2}$. Therefore $\mathcal{M}%
_{\mathbf{k}\mathbf{q}}\simeq \sqrt{\frac{\hbar c_{\text{b}}k}{2g_{\text{b}%
}n_{\text{b}}}}$. Moreover, in this regime we can approximate $\mathcal{E}%
_{B_{\text{b}}}(\mathbf{k})=\hbar c_{\text{b}}|\mathbf{k}|$ and $\mathcal{E}%
_{F_{\text{f}}}(\mathbf{k})=\Delta +\frac{\hbar ^{2}}{2m^{\ast }}\eta _{%
\mathbf{k}}^{2}$, where $\eta _{\mathbf{k}}=|\mathbf{k}|-k_{0}$, $m^{\ast
}=\Delta /v_{0}^{2}$ and $v_{0}=\hbar k_{0}/m_{\text{f}}$. By this
approximation, $k_{\text{c}}$ is determined by $2\Delta =\hbar c_{\text{b}%
}k_{\text{c}}$. With these approximations, the damping rate $\gamma (\mathbf{%
k})$ can be simplified as
\begin{equation}
\gamma (\mathbf{k})=\frac{g_{\text{bf}}^{2}c_{\text{b}}k}{8\pi ^{2}g_{\text{b%
}}}\int d^{3}\mathbf{q}\delta \left[ \frac{\hbar ^{2}\left( \eta _{\mathbf{%
k-q}}^{2}+\eta _{\mathbf{q}}^{2}\right) }{2m^{\ast }}-\hbar c_{\text{b}%
}(k-k_{\text{c}})\right] .
\end{equation}%
Basically this integration is to count for the density-of-state that
satisfies energy conservation. With quite straightforward calculation \cite%
{supple} we find that
\begin{equation}
\gamma (\mathbf{k})=\frac{g_{\text{bf}}^{2}c_{\text{b}}\Delta m_{\text{f}%
}^{2}}{2\hbar ^{4}g_{\text{b}}}\Theta (k-k_{\text{c}}),
\end{equation}%
i.e. the threshold behavior of $\gamma (\mathbf{k})$ is a constant.

\begin{figure}[t]
\includegraphics[width=3.5 in]
{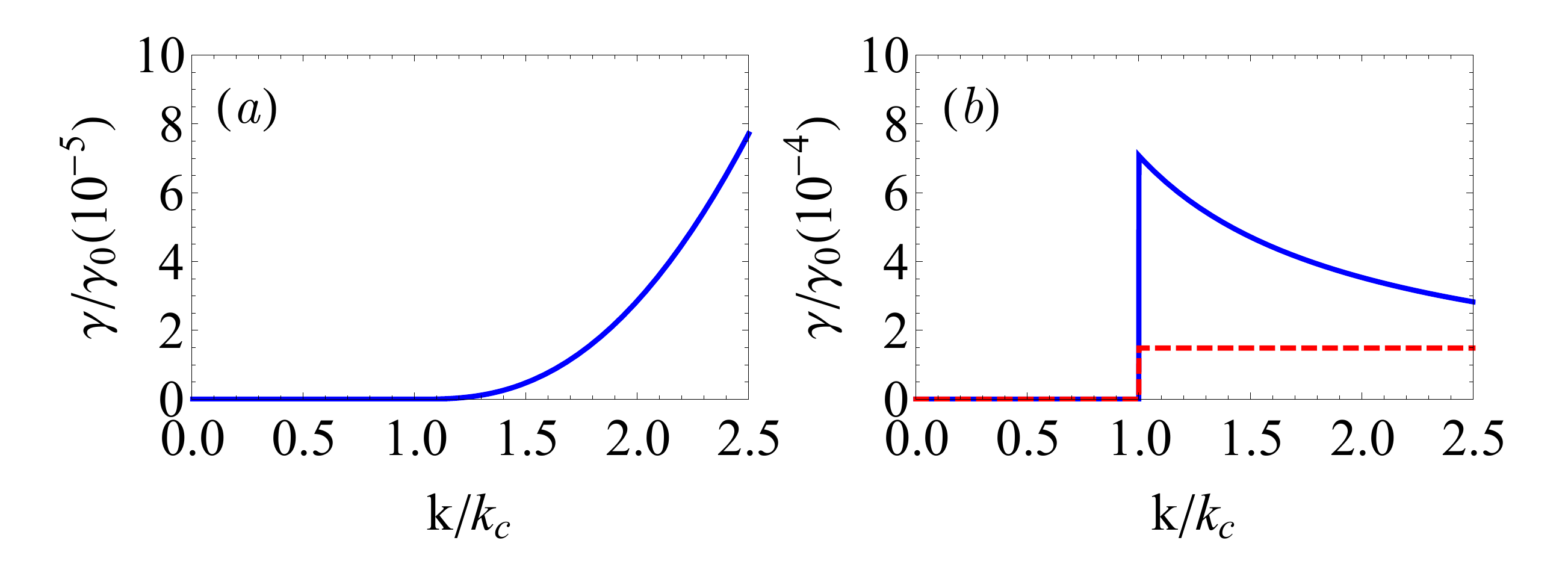}
\caption{Damping rate $\protect\gamma $ in unit of $\protect\gamma _{0}$ ($%
\protect\gamma _{0}=E_{\text{F}}/\hbar $) as a function of $k/k_{\text{c}}$.
(a): Case A and B in the BCS side are shown with $1/(k_{\text{F}}a_{\text{f}%
})=-0.25$ for Case A (dashed line) and $1/(k_{\text{F}}a_{\text{f}})=-0.5$
for Case B (solid line). (b) Case C in the BEC side with $1/(k_{\text{F}}a_{%
\text{f}})=0.5$. For a typical experiment setup $k_{\text{F}}\approx 5\times
10^{6}\mathrm{m}^{-1}$ and $\protect\gamma _{0}\approx 1.3\times 10^{5}%
\mathrm{Hz}$.}
\label{damping}
\end{figure}

\textit{Case B.} In this regime the damping rate is still determined by Eq. %
\ref{rate_BCS}. But since $k_{\text{c}}$ is in the free-particle regime, we
have $u_{\mathbf{k}}^{\text{b}}\approx 1$ and $v_{\mathbf{k}}^{\text{b}%
}\approx 0$. In this case $\mathcal{M}_{\mathbf{k}\mathbf{q}}\approx 1$.
Furthermore, $\mathcal{E}_{B_{\text{b}}}(\mathbf{k})$ is approximated by $%
\epsilon _{\text{b}}(\mathbf{k})+g_{\text{b}}n_{\text{b}}$, and the damping
rate $\gamma (\mathbf{k})$ is given by
\begin{equation}
\gamma (\mathbf{k})=\frac{g_{\text{bf}}^{2}n_{\text{b}}}{4\pi ^{2}\hbar }%
\int d^{3}\mathbf{q}\delta \left[ \frac{\hbar ^{2}\left( \eta _{\mathbf{k-q}%
}^{2}+\eta _{\mathbf{q}}^{2}\right) }{2m^{\ast }}-\frac{\hbar ^{2}\left(
\mathbf{k}^{2}-k_{\text{c}}^{2}\right) }{2m_{\text{b}}}\right] ,
\end{equation}%
which gives rise to a damping rate
\begin{align}
\gamma (\mathbf{k})& =\frac{g_{\text{bf}}^{2}n_{\text{b}}\Delta m_{\text{f}%
}^{2}}{\hbar ^{5}k}\Theta (k-k_{\text{c}})  \notag \\
& \simeq \frac{g_{\text{bf}}^{2}n_{\text{b}}\Delta m_{\text{f}}^{2}}{\hbar
^{5}k_{\text{c}}}\left( 1-\frac{k-k_{\text{c}}}{k_{\text{c}}}\right) \Theta
(k-k_{\text{c}}).
\end{align}%
The leading order is still a constant and the sub-leading order gives a slow
decreasing of $\gamma (\mathbf{k})$ as $|\mathbf{k}|$ increases. However, we
shall also note that because the approximations implemented, our results are
only valid nearby $k_{\text{c}}$ and cannot be extended to very large
momentum.

\textit{Case C.} In this regime the damping is due to coupling between $B_{%
\text{b}}$ mode and $B_{\text{f}}$ mode. A comprehensive description of $B_{%
\text{f}}$ mode and its coupling to $B_{\text{b}}$ mode can be obtained from
fluctuation theory of Fermi superfluid \cite{bmode}. Here to highlight the
essential physics we take a simpler approach by treating the Fermi
superfluid at the BEC side as molecular condensate, and we consider a
Hamiltonian of molecular BEC as
\begin{equation}
\hat{H}_{\text{m}}=\int d^{3}\mathbf{r}\left\{ \hat{d}^{\dag }(\mathbf{r}%
)H_{0,\text{m}}\hat{d}(\mathbf{r})+\frac{g_{\text{m}}}{2}\hat{d}^{\dag }(%
\mathbf{r})\hat{d}^{\dag }(\mathbf{r})\hat{d}(\mathbf{r})\hat{d}(\mathbf{r}%
)\right\}
\end{equation}%
where $H_{0,\text{m}}=-\frac{\hbar ^{2}\nabla ^{2}}{2m_{\text{m}}}-\mu _{%
\text{m}}$, and $g_{\text{m}}=4\pi \hbar ^{2}a_{\text{m}}/m_{\text{m}}$. $%
\hat{d}^\dag$ represents a creation operator for a bosonic molecule. The
coupling between the molecular BEC and Bose superfluid is due to scattering
between bosonic atoms and molecules, which can be effectively described by
\begin{equation}
H_{\text{bm}}=g_{\text{bm}}\int d^{3}\mathbf{r}\hat{b}^{\dag }(\mathbf{r})%
\hat{b}(\mathbf{r})\hat{d}^{\dag }(\mathbf{r})\hat{d}(\mathbf{r})
\end{equation}%
where $g_{\text{bm}}$ is determined by atom-molecule scattering length
calculated in Ref. \cite{ad}. Bogoliubov approximation can be applied to $%
\hat{H}_{\text{m}}$ which gives
\begin{equation}
\hat{H}_{\text{m}}=\sum\limits_{\mathbf{k}}\mathcal{E}_{B_{\text{f}}}(%
\mathbf{k})\hat{\chi}_{\mathbf{k}}^{\dag }\hat{\chi}_{\mathbf{k}},
\end{equation}%
where $\mathcal{E}_{B_{\text{f}}}(\mathbf{k})=\sqrt{\epsilon _{\text{m}}(%
\mathbf{k})\left[ \epsilon _{\text{m}}(\mathbf{k})+2g_{\text{m}}n_{\text{m}}%
\right] }$ with $\epsilon _{\text{m}}(\mathbf{k})=\hbar ^{2}\mathbf{k}%
^{2}/(2m_{\text{m}})$. $\hat{\chi}_{\mathbf{k}}$ relates to $\hat{d}_{%
\mathbf{k}}$ as $\hat{\chi}_{\mathbf{k}}=u_{\mathbf{k}}^{\text{m}}\hat{d}_{%
\mathbf{k}}-v_{\mathbf{k}}^{\text{m}}\hat{d}_{\mathbf{-k}}^{\dag }$, where $%
u_{\mathbf{k}}^{\text{m}}(v_{\mathbf{k}}^{\text{m}})=\sqrt{\frac{1}{2}\left(
\frac{\epsilon _{\text{m}}(\mathbf{k})+g_{\text{m}}n_{\text{m}}}{\mathcal{E}%
_{B_{\text{f}}}(\mathbf{k})}\pm 1\right) }$. Similarly, in the order
proportional to $n_{\text{b}}$ or $n_{\text{m}}$, $\hat{H}_{\text{bm}}$ is
simply a constant chemical potential shift for both Bose superfluid and
molecular condensate.

Similar as analysis in case A, by replacing one of $\hat{d}^{\dag }$ (or $%
\hat{d}$) operator as $\sqrt{N_{\text{m}}}$ or one of $\hat{b}^{\dag }$ (or $%
\hat{b}$) operator as $\sqrt{N_{\text{b}}}$, it can be expanded into quite a
few terms that describe quasi-particle interactions, among which only one
term contributes to decay of $B_{\text{b}}$ mode with a lower critical
velocity, as discussed above \cite{supple}. This term is given by
\begin{align}
& g_{\text{bm}}\sqrt{\frac{n_{\text{m}}}{V}}\sum\limits_{\mathbf{k}\mathbf{q}%
}\mathcal{Q}_{\mathbf{k}\mathbf{q}}\hat{\chi}_{\mathbf{q}}^{\dag }\hat{\alpha%
}_{\mathbf{k-q}}^{\dag }\hat{\alpha}_{\mathbf{k}}, \\
& \mathcal{Q}_{\mathbf{k}\mathbf{q}}=(u_{\mathbf{q}}^{\text{m}}-v_{\mathbf{q}%
}^{\text{m}})(u_{\mathbf{k-q}}^{\text{b}}u_{\mathbf{k}}^{\text{b}}+v_{%
\mathbf{k-q}}^{\text{b}}v_{\mathbf{k}}^{\text{b}}).
\end{align}%
In this regime we can approximate $u_{\mathbf{k}}^{\text{m}}(v_{\mathbf{k}}^{%
\text{m}})=\sqrt{\frac{g_{\text{m}}n_{\text{m}}}{2\hbar c_{\text{f}}k}}\pm
\frac{1}{2}\sqrt{\frac{\hbar c_{\text{f}}k}{2g_{\text{m}}n_{\text{m}}}}$, $%
u_{\mathbf{k}}^{\text{b}}\approx 1$ and $v_{\mathbf{k}}^{\text{b}}\approx 0$%
, therefore $\mathcal{Q}_{\mathbf{k}\mathbf{q}}$ becomes $\sqrt{\frac{\hbar
c_{\text{f}}k}{2g_{\text{m}}n_{\text{m}}}}$. Furthermore, we can approximate
$\mathcal{E}_{B_{\text{b}}}(\mathbf{k})$ by $\epsilon _{\text{b}}(\mathbf{k}%
)+g_{\text{b}}n_{\text{b}}$, $\mathcal{E}_{B_{\text{f}}}$ as $\hbar c_{\text{%
f}}|\mathbf{k}|$, and the damping rate is
\begin{equation}
\gamma (\mathbf{k})=\frac{g_{\text{bm}}^{2}c_{\text{f}}}{8\pi ^{2}g_{\text{m}%
}}\int d^{3}\mathbf{q}|\mathbf{q}|\delta \left\{ \hbar c_{\text{f}}|\mathbf{q%
}|+\frac{\hbar ^{2}\left[ (\mathbf{k-q})^{2}-\mathbf{k}^{2}\right] }{2m_{%
\text{b}}}\right\} .
\end{equation}%
Straightforward evaluation of this integral gives \cite{supple}
\begin{equation}
\gamma (\mathbf{k})=\frac{2g_{\text{bm}}^{2}m_{\text{b}}c_{\text{f}}}{3\pi
\hbar ^{2}g_{\text{m}}k}(k-k_{\text{c}})^{3}\Theta (k-k_{\text{c}}).
\end{equation}%
At leading order $\gamma (\mathbf{k})$ fast increases as $(k-k_{\text{c}%
})^{3}$ once $k$ is above threshold.

\textit{Conclusion.} The results of damping rate for three cases are
presented in Fig. \ref{damping}. We choose $n_{\text{b}}/k_{\text{F}%
}^{3}=0.1 $, $1/\left( k_{\text{F}}a_{\text{b}}\right) =100$ and $1/\left(
k_{\text{F}}a_{\text{bf}}\right) =100$. For three different cases, we choose
$1/\left( k_{\text{F}}a_{\text{f}}\right) =-2.5$, $1/\left( k_{\text{F}}a_{%
\text{f}}\right) =-0.5$, and $1/\left( k_{\text{F}}a_{\text{f}}\right) =0.5$%
, respectively. We find a different threshold behavior $\gamma (\mathbf{k}%
)\propto (k-k_{\text{c}})^{\alpha }$ with $\alpha =0$ in the BCS regime and $%
\alpha =3$ in the BEC regime. This finding, on one hand, is a unique
manifestation of quasi-particle interaction effect in the Bose-Fermi
superfluid mixture; on the other hand, reveals fundamental different between
Fermi superfluid in the BCS side and in the BEC side. In the BCS side, the
low-energy physics is dominated by fermionic quasi-particles nearby the
Fermi surface, and the damping processes are also restricted by the constant
density-of-state nearby Fermi surface, which is basically the origin of
constant damping rate. While such restriction does not exist in the BEC side
where the low-energy physics is dominated by bosonic mode.

Our results can be experimentally verified by studying damping rate of
collective mode, as done in previous BEC experiments \cite{exp}. In the
recent experiment, ENS group has find damping of collective oscillation when
the relative velocity between Bose and Fermi superfluid exceeds a critical
velocity. At the unitary regime and in the BEC side the damping rate
increases rapidly when velocity is above the critical velocity \cite{ENS}.
They also find a nearly constant damping rate at the BCS side \cite{private}%
. The underlying mechanism of this experimental finding may be connected to
the physics discussed in this work.

\textit{Acknowledgment}: We wish to thank Christophe Salomon for helpful
discussions. This work is supported by Tsinghua University Initiative
Scientific Research Program, NSFC Grant No. 11174176, and NKBRSFC under
Grant No. 2011CB921500.

\begin{widetext}
\section{ Supplemental material }

\subsection{The damping rate in the BCS side}

From Eq. (1) in the text, the Hamiltonian describing the interaction between
the Bosons and Fermions in momentum space is given by%
\begin{equation}
H_{\text{bf}}=\frac{g_{\text{bf}}}{V}\sum\limits_{\mathbf{q,p,k,}\sigma }%
\hat{c}_{\mathbf{q-k,}\sigma }^{\dag }\hat{c}_{\mathbf{q,}\sigma }\hat{b}_{%
\mathbf{p+k}}^{\dag }\hat{b}_{\mathbf{p}}.
\end{equation}%
By replacing one of the $\hat{b}$ or $\hat{b}^{\dag }$ operators as $\sqrt{%
N_{\text{b}}}$, this Hamiltonian becomes%
\begin{equation}
H_{\text{bf}}=g_{\text{bf}}\sqrt{\frac{n_{\text{b}}}{V}}\sum\limits_{\mathbf{%
k,q,}\sigma }\left( \hat{c}_{\mathbf{k+q,}\sigma }^{\dag }\hat{c}_{\mathbf{q,%
}\sigma }\hat{b}_{\mathbf{k}}+\text{h.c.}\right) .
\end{equation}%
Then we rewrite the Hamiltonian in terms of the quasi-particle operators as%
\begin{eqnarray}
H_{\text{bf}} &=&g_{\text{bf}}\sqrt{\frac{n_{\text{b}}}{V}}\sum\limits_{%
\mathbf{k,q}}\left( u_{\mathbf{k}}^{\text{b}}-v_{\mathbf{k}}^{\text{b}%
}\right) \left( u_{\mathbf{k-q}}^{\text{f}}v_{\mathbf{k}}^{\text{f}}+v_{%
\mathbf{k-q}}^{\text{f}}u_{\mathbf{k}}^{\text{f}}\right) \hat{\beta}_{%
\mathbf{k-q}}^{\dag }\hat{\gamma}_{\mathbf{q}}^{\dag }\hat{\alpha}_{\mathbf{k%
}}+\text{h.c.}  \notag \\
&&+g_{\text{bf}}\sqrt{\frac{n_{\text{b}}}{V}}\sum\limits_{\mathbf{k,q}%
}\left( u_{\mathbf{k}}^{\text{b}}-v_{\mathbf{k}}^{\text{b}}\right) \left( u_{%
\mathbf{k+q}}^{\text{f}}u_{\mathbf{k}}^{\text{f}}-v_{\mathbf{k+q}}^{\text{f}%
}v_{\mathbf{k}}^{\text{f}}\right) \hat{\beta}_{\mathbf{k+q}}^{\dag }\hat{%
\beta}_{\mathbf{q}}\hat{\alpha}_{\mathbf{k}}+\text{h.c.}  \notag \\
&&+g_{\text{bf}}\sqrt{\frac{n_{\text{b}}}{V}}\sum\limits_{\mathbf{k,q}%
}\left( u_{\mathbf{k}}^{\text{b}}-v_{\mathbf{k}}^{\text{b}}\right) \left( u_{%
\mathbf{k+q}}^{\text{f}}u_{\mathbf{k}}^{\text{f}}-v_{\mathbf{k+q}}^{\text{f}%
}v_{\mathbf{k}}^{\text{f}}\right) \hat{\gamma}_{\mathbf{k+q}}^{\dag }\hat{%
\gamma}_{\mathbf{q}}\hat{\alpha}_{\mathbf{k}}+\text{h.c.}.
\end{eqnarray}%
Here we have ignored the terms such as $\hat{\beta}\hat{\gamma}\hat{\alpha}$
or $\hat{\beta}^{\dag }\hat{\gamma}^{\dag }\hat{\alpha}^{\dag }$, since they
do not conserve the energy, and will not contribute to the decay process. At
the zero temperature, only the first term contribute to the damping of the $%
B_{\text{b}}$ mode.

Employing the approximations discussed in the text, we obtain the integral
as Eq. (8) in the text. To calculate this integral, we first make the
substitution: $\mathbf{k-q}\rightarrow \frac{\mathbf{k}}{2}-\mathbf{q}$ and $%
\mathbf{q}\rightarrow \frac{\mathbf{k}}{2}+\mathbf{q}$, so that the integral
becomes%
\begin{equation}
\gamma \left( \mathbf{k}\right) =\frac{g_{\text{bf}}^{2}c_{\text{b}}k}{8\pi
^{2}g_{\text{b}}}\int d^{3}\mathbf{q}\delta \left[ \frac{\hbar ^{2}}{%
2m^{\ast }}\left( \eta _{\frac{\mathbf{k}}{2}-\mathbf{q}}^{2}+\eta _{\frac{%
\mathbf{k}}{2}+\mathbf{q}}^{2}\right) -\hbar c_{\text{b}}\left( k-k_{\text{c}%
}\right) \right] .
\end{equation}%
We choose the direction of $\mathbf{k}$ as the $q_{z}$ axis, and transform
into the cylindrical polar coordinates. The coordinate transformation is
given by%
\begin{eqnarray*}
q_{z} &=&p_{z}, \\
q_{x} &=&\left( k_{0}\sin \theta _{0}+p_{\rho }\right) \cos \phi , \\
q_{y} &=&\left( k_{0}\sin \theta _{0}+p_{\rho }\right) \sin \phi .
\end{eqnarray*}%
where $\theta _{0}$ is defined as $\cos \theta _{0}=\frac{k}{2k_{0}}$. The
Jacobi determinant of this coordinate transformation is $dq_{z}dq_{x}dq_{y}=%
\left( k_{0}\sin \theta _{0}+p_{\rho }\right) dp_{z}dp_{\rho }d\phi $. Then
the $\eta _{\frac{\mathbf{k}}{2}\pm \mathbf{q}}$ can be expanded in the new
coordinates as%
\begin{equation*}
\eta _{\frac{\mathbf{k}}{2}\pm \mathbf{q}}=\left\vert \frac{\mathbf{k}}{2}%
\pm \mathbf{q}\right\vert -k_{0}\approx \sin \theta _{0}p_{\rho }\pm \cos
\theta _{0}p_{z},
\end{equation*}%
where the high order terms of $p_{\rho }$ and $p_{z}$ are ignored. Then the
integral becomes%
\begin{eqnarray}
\gamma \left( \mathbf{k}\right)  &=&\frac{g_{\text{bf}}^{2}c_{\text{b}}k}{%
8\pi ^{2}g_{\text{b}}}\int_{-\infty }^{\infty }dp_{z}\int_{-k_{0}\sin \theta
_{0}}^{\infty }dp_{\rho }\int_{0}^{2\pi }d\phi   \notag \\
&&\times \left( k_{0}\sin \theta _{0}+p_{\rho }\right) \delta \left[ \frac{%
\hbar ^{2}}{2m^{\ast }}\left( p_{\rho }^{2}\sin ^{2}\theta
_{0}+p_{z}^{2}\cos ^{2}\theta _{0}\right) -\hbar c_{\text{b}}\left( k-k_{%
\text{c}}\right) \right] .  \label{Int1}
\end{eqnarray}%
We apply a coordinate transformation again as%
\begin{eqnarray*}
-\frac{\hbar }{\sqrt{m^{\ast }}}p_{\rho }\sin \theta _{0} &=&r\cos \zeta , \\
\frac{\hbar }{\sqrt{m^{\ast }}}p_{z}\cos \theta _{0} &=&r\sin \zeta .
\end{eqnarray*}%
where the corresponding Jacobi determinant is $dp_{z}dp_{\rho }=\frac{%
m^{\ast }}{\hbar ^{2}\sin \theta _{0}\cos \theta _{0}}rdrd\zeta $. So Eq.\ref%
{Int1} becomes%
\begin{eqnarray}
\gamma \left( \mathbf{k}\right)  &=&\frac{g_{\text{bf}}^{2}c_{\text{b}}k}{%
4\pi g_{\text{b}}}\int_{0}^{\infty }dr\int_{\zeta _{0}}^{2\pi -\zeta
_{0}}d\zeta \frac{m^{\ast }}{\hbar ^{2}\sin \theta _{0}\cos \theta _{0}}r
\notag \\
&&\times \left( k_{0}\sin \theta _{0}-\frac{\sqrt{m^{\ast }}}{\hbar \sin
\theta _{0}}r\cos \zeta \right) \delta \left[ r^{2}-\hbar c_{\text{b}}\left(
k-k_{\text{c}}\right) \right] .
\end{eqnarray}%
where $\zeta _{0}$ is given by $\left\vert \cos \zeta _{0}\right\vert =\frac{%
\hbar k_{0}\sin \theta _{0}}{\sqrt{2m^{\ast }\hbar c_{\text{b}}\left( k-k_{%
\text{c}}\right) }}$. Since we are Considering the threshold behavior, we
have $\zeta _{0}=0$. The damping rate is obtained as%
\begin{equation}
\gamma \left( \mathbf{k}\right) =\frac{g_{\text{bf}}^{2}c_{\text{b}%
}k_{0}^{2}m^{\ast }}{2\hbar ^{2}g_{\text{b}}}\Theta \left( k-k_{\text{c}%
}\right)
\end{equation}%
Substituting the expression of the effective mass, $m^{\ast }=\frac{\Delta
m_{\text{f}}^{2}}{\hbar ^{2}k_{0}^{2}}$, to the upper formula, one obtains
the Eq. (9) in the text.

In the free boson regime, $\Delta \gg \hbar ^{2}/\left( m_{\text{b}}\xi
^{2}\right) $, using the same integral skill, one obtains the damping rate
as:%
\begin{equation}
\gamma \left( \mathbf{k}\right) =\frac{g_{\text{bf}}^{2}n_{\text{b}%
}k_{0}^{2}m^{\ast }}{\hbar ^{3}k}\Theta \left( k-k_{\text{c}}\right)
\end{equation}%
Substituting the expression of the effective mass to the upper formula, we
have the Eq. (11) in the text.

\subsection{Damping rate in the BEC side}

From Eq. (12), in the BEC side the Boson-molecular interaction Hamiltonian
in the momentum space is given by%
\begin{equation}
H_{\text{bm}}=\frac{g_{\text{bm}}}{V}\sum\limits_{\mathbf{q,p,k}}\hat{d}_{%
\mathbf{q-k}}^{\dag }\hat{d}_{\mathbf{q}}\hat{b}_{\mathbf{p+k}}^{\dag }\hat{b%
}_{\mathbf{p}}.  \label{Hbm}
\end{equation}%
By replacing one of the $\hat{d}$ or $\hat{d}^{\dag }$ operators by the $%
\sqrt{N_{\text{m}}}$, we obtain%
\begin{equation}
H_{\text{bm}}^{(1)}=g_{\text{bm}}\sqrt{\frac{n_{\text{m}}}{V}}\sum\limits_{%
\mathbf{k,q}}\left( \hat{d}_{-\mathbf{q}}+\hat{d}_{\mathbf{q}}^{\dag
}\right) \hat{b}_{\mathbf{k-q}}^{\dag }\hat{b}_{\mathbf{k}}.
\end{equation}%
This term describes the $B_{\text{b}}$ mode scattered by the phonon mode $B_{%
\text{f}}$ in the Fermi superfluid. We rewrite this Hamiltonian in terms of
the quasi-particle operators%
\begin{eqnarray}
H_{\text{bm}}^{(1)} &=&g_{\text{bm}}\sqrt{\frac{n_{\text{m}}}{V}}%
\sum\limits_{\mathbf{k,q}}\left( u_{\mathbf{q}}^{\text{m}}-v_{\mathbf{q}}^{%
\text{m}}\right) \left( u_{\mathbf{k-q}}^{\text{b}}u_{\mathbf{k}}^{\text{b}%
}+v_{\mathbf{k-q}}^{\text{b}}v_{\mathbf{k}}^{\text{b}}\right) \hat{\chi}_{%
\mathbf{q}}^{\dag }\hat{\alpha}_{\mathbf{k-q}}^{\dag }\hat{\alpha}_{\mathbf{k%
}}+\text{h.c.}  \notag \\
&&-g_{\text{bm}}\sqrt{\frac{n_{\text{m}}}{V}}\sum\limits_{\mathbf{k,q}%
}\left( u_{\mathbf{q}}^{\text{m}}-v_{\mathbf{q}}^{\text{m}}\right) v_{%
\mathbf{q-k}}^{\text{b}}u_{\mathbf{k}}^{\text{b}}\hat{\chi}_{\mathbf{q}%
}^{\dag }\hat{\alpha}_{\mathbf{q-k}}\hat{\alpha}_{\mathbf{k}}+\text{h.c.},
\end{eqnarray}%
where such terms as $\hat{\chi}\hat{\alpha}\hat{\alpha}$ or $\hat{\chi}%
^{\dag }\hat{\alpha}^{\dag }\hat{\alpha}^{\dag }$ are ignored, since they do
not conserve the energy. At the zero temperature, only the first term
contribute to the damping, which describes the process $B_{\text{b}}(\mathbf{%
k})\rightarrow B_{\text{b}}(\mathbf{k-q})+B_{\text{f}}(\mathbf{q})$. The
critical momentum of this process can be obtained by energy-momentum
conservation as $\left. \partial \mathcal{E}_{B_{\text{b}}}(k)/\partial
\left( \hbar k\right) \right\vert _{k=k_{\text{c}}}=c_{\text{f}}$. In the
free boson regime, we have $\hbar k_{\text{c}}=m_{\text{f}}c_{\text{f}}$.

By replacing one of the $\hat{b}^{\dag }$ or $\hat{b}$ operators by the $%
\sqrt{N_{\text{b}}}$ in Eq. \ref{Hbm}, we obtain%
\begin{eqnarray}
H_{\text{bm}}^{(2)} &=&-g_{\text{bm}}\sqrt{\frac{n_{\text{b}}}{V}}%
\sum\limits_{\mathbf{k,q}}\left( u_{\mathbf{k}}^{\text{b}}-v_{\mathbf{k}}^{%
\text{b}}\right) u_{\mathbf{k-q}}^{\text{m}}v_{\mathbf{q}}^{\text{m}}\hat{%
\chi}_{\mathbf{k-q}}^{\dag }\hat{\chi}_{\mathbf{q}}^{\dag }\hat{\alpha}_{%
\mathbf{k}}+\text{h.c.}  \notag \\
&&+g_{\text{bm}}\sqrt{\frac{n_{\text{b}}}{V}}\sum\limits_{\mathbf{k,q}%
}\left( u_{\mathbf{k}}^{\text{b}}-v_{\mathbf{k}}^{\text{b}}\right) \left( u_{%
\mathbf{k+q}}^{\text{m}}u_{\mathbf{q}}^{\text{m}}+v_{\mathbf{k+q}}^{\text{m}%
}v_{\mathbf{q}}^{\text{m}}\right) \hat{\chi}_{\mathbf{k+q}}^{\dag }\hat{\chi}%
_{\mathbf{q}}\hat{\alpha}_{\mathbf{k}}+\text{h.c.},
\end{eqnarray}%
At zero temperature, only the first term contribute to the damping, which
describes the process $B_{\text{b}}(\mathbf{k})\rightarrow B_{\text{f}}(%
\mathbf{k-q})+B_{\text{f}}(\mathbf{q})$. Using the energy-momentum
conservation, one can determine the critical momentum for this process by $%
\mathcal{E}_{B_{\text{b}}}(k_{\text{c}})=2\mathcal{E}_{B_{\text{f}}}\left(
k_{\text{c}}/2\right) $. In the free boson regime, we have $\hbar k_{\text{c}%
}\approx 2m_{\text{f}}c_{\text{f}}$, which is larger than the critical
momentum of the process discussed above. So we will focus on the threshold
behavior of the process $B_{\text{b}}\rightarrow B_{\text{b}}+B_{\text{f}}$.

Employing the approximations discussed in the text, we obtain damping rate
as Eq. (17) in the text:%
\begin{equation}
\gamma \left( \mathbf{k}\right) =\frac{g_{\text{bm}}^{2}c_{\text{f}}}{8\pi
^{2}g_{\text{m}}}\int d^{3}\mathbf{q}\delta \left[ \hbar c_{\text{f}%
}\left\vert \mathbf{q}\right\vert +\frac{\hbar ^{2}\left( \mathbf{k-q}%
\right) ^{2}}{2m_{\text{b}}}-\frac{\hbar ^{2}\mathbf{k}^{2}}{2m_{\text{b}}}%
\right] \left\vert \mathbf{q}\right\vert .
\end{equation}%
The Dirac function in this integral gives%
\begin{equation}
\delta \left[ \hbar c_{\text{f}}\left\vert \mathbf{q}\right\vert +\frac{%
\hbar ^{2}\left( \mathbf{k-q}\right) ^{2}}{2m_{\text{b}}}-\frac{\hbar ^{2}%
\mathbf{k}^{2}}{2m_{\text{b}}}\right] =\frac{m_{\text{b}}}{\hbar ^{2}kq}%
\delta \left( \frac{\hbar q+2m_{\text{b}}c_{\text{f}}}{2\hbar k}-\cos \theta
\right) ,
\end{equation}%
where $\theta $ is the angle between $\mathbf{q}$ and $\mathbf{k}$. So the
integral becomes%
\begin{equation}
\gamma \left( \mathbf{k}\right) =\frac{g_{\text{bm}}^{2}m_{\text{b}}c_{\text{%
f}}}{4\pi \hbar ^{2}g_{\text{m}}k}\int_{0}^{\infty }q^{2}dq\int_{\pi
}^{0}d\cos \theta \delta \left( \frac{\hbar q+2m_{\text{b}}c_{\text{f}}}{%
2\hbar k}-\cos \theta \right) .
\end{equation}%
Since we have $-1<\cos \theta <1$, the integral regime of $q$ is determined
as%
\begin{equation}
-2\left( k+\frac{m_{\text{b}}c_{\text{f}}}{\hbar }\right) <q<2\left( k-\frac{%
m_{\text{b}}c_{\text{f}}}{\hbar }\right) .
\end{equation}%
The integral can be simplified into%
\begin{eqnarray}
\gamma \left( \mathbf{k}\right) &=&\frac{g_{\text{bm}}^{2}m_{\text{b}}c_{%
\text{f}}}{4\pi \hbar ^{2}g_{\text{m}}k}\int_{0}^{2\left( k-m_{\text{b}}c_{%
\text{f}}/\hbar \right) }q^{2}dq.  \notag \\
&=&\frac{2g_{\text{bm}}^{2}m_{\text{b}}c_{\text{f}}}{3\pi \hbar ^{2}g_{\text{%
m}}k}\left( k-m_{\text{b}}c_{\text{f}}/\hbar \right) ^{3}\Theta \left( k-m_{%
\text{b}}c_{\text{f}}/\hbar \right) .  \notag \\
&=&\frac{2g_{\text{bm}}^{2}m_{\text{b}}c_{\text{f}}}{3\pi \hbar ^{2}g_{\text{%
m}}k}\left( k-k_{\text{c}}\right) ^{3}\Theta \left( k-k_{\text{c}}\right) .
\end{eqnarray}%
Here we can see $k_{\text{c}}=m_{\text{b}}c_{\text{f}}/\hbar $, which
reproduces the critical momentum discussed above.

\end{widetext}

\end{document}